\documentclass[aps,floatfix,pra,reprint,superscriptaddress,nofootinbib]{revtex4-2}

\usepackage{amsmath,amssymb,bm}
\usepackage{graphicx}
\usepackage{dcolumn}   
\usepackage{hyperref}  
\usepackage{braket}
\usepackage{comment}
\usepackage{cancel}
\hypersetup{colorlinks=true,allcolors=blue}

\begin{document}

\title{Low energy elastic scattering of hydrogen, deuterium and tritium on helium isotopes}
\author{B.J.P. Jones}
  \email{benjamin.jones-7@manchester.ac.uk}
  \affiliation{Department of Physics and Astronomy, University of Manchester, Oxford Road, Manchester, M13 9PY, United Kingdom}
  \affiliation{Department of Physics, University of Texas at Arlington, Arlington, TX 76019, USA}

 \author{A. Negi}
  \affiliation{Department of Physics, University of Texas at Arlington, Arlington, TX 76019, USA}

  \author{A. Semakin}
  \affiliation{Wihuri Physical Laboratory, Department of Physics, University of Turku, 20014 Turku, Finland}
\date{\today}

\begin{abstract}
   Motivated by the needs of atomic tritium sources for neutrino mass experiments and Doppler-free two-photon 1S–2S spectroscopy in atomic deuterium and tritium, we present calculations of energy-dependent elastic scattering cross sections of hydrogen isotopes (H, D and T) on helium isotopes ($^3$He and $^4$He) in the temperature range 1~mK to 300~K.  The tritium-on-helium cross sections are found to be enhanced over their hydrogen-on-helium counterparts by a near-threshold resonant \textit{s}-wave bound state at low energy, similar to one that has been predicted in the triplet T-T system. While the energy-dependent cross sections span a wide range at low energy due to this \textit{s}-wave enhancement, they tend toward a common value at high energy where the scattering becomes effectively geometric in nature.  
\end{abstract}

\maketitle
\section{Introduction}

The low energy elastic scattering of hydrogen-like atoms on each other
and on isotopes of helium is of interest in a variety of contexts
in atomic physics. In particular, the dynamics of cold atomic tritium
are receiving continuing attention in the context of experiments
that aim to measure the mass of the neutrino~\cite{formaggio2021direct} using atomic~(T) rather
than molecular (T$_{2}$) sources~\cite{pettus2020overview}. To use T for this purpose, molecular
tritium must be dissociated~\cite{outten1991characterization,tschersich1998formation} and then cooled to the point where it
can be stored magnetically in large volumes,  which is feasible at temperatures of at
most a few tens of milikelvin. The vapor properties, evaporative cooling
efficiency and loss rates of tritium atoms in proposed cooling
systems and magnetic traps depends strongly on their elastic and inelastic
scattering cross sections~\cite{esfahani2025dynamics}, which are as-yet unmeasured.

To facilitate the designs of atomic T cooling systems for proposed neutrino mass
experiments including Project~8~\cite{ashtari2023tritium}, KATRIN++~\cite{kovac2025katrinpp} and QTNM~\cite{amad2025determining}, and cryogenic dissociation sources~\cite{semakin2025cryogenic}, some of us recently
produced a new suite of calculations of H-H and T-T scattering in
all elastic and inelastic channels as a function of magnetic field
and temperature~\cite{ElliottJones2025}. These predictions enable advances in the quantitative design and optimization
of properties of atomic tritium cooling systems~\cite{esfahani2025dynamics}.  

This paper presents a suite of calculations for a second set of relevant processes, obtained using a subset of the methods of Ref~\cite{ElliottJones2025}.  These are the cross sections of atomic hydrogen isotopes
scattering from helium atoms.  While less central to experiment design than their T-T scattering counterparts, tritium-helium scattering informs the designs of atomic tritium sources for neutrino physics in at least the
following contexts: 
\begin{enumerate}
\item The decay of T to $^{3}$He occurs continuously within an atomic T
source, and hence trace levels of $^{3}$He are both initially present in incoming T$_2$ gas and produced continuously within trapped T vapor. Since $^{3}$He
is not itself magnetically trapped it will ultimately be either turbo- or
cryo-pumped from the active volume. Nevertheless, a stable background level of
$^{3}$He (most of which is thermalised with the vessel walls) is expected in the tritium cell, leading to T-$^3$He
scattering, which should be accounted for when modeling tritium vapor dynamics. The low energy elastic scattering cross section of T on $^3$He has not been previously reported.

\item Slowing of tritium on cryogenic buffer gas~\cite{hutzler2012buffer} has been discussed as a possible way
to partially slow and cool atomic T. As the lightest gases which
have no chemical activity, $^{3}$He and $^{4}$He are  natural
choices for this purpose. The slowing and cooling rates are determined
by the energy dependent T-$^{3}$He and T-$^{4}$He cross sections.
\item Supersonic expansion sources have been proposed to produce cold beams
of H and T~\cite{amad2025determining}.  One possible approach for atomic tritium beam production, appealing because it avoids handling compressed tritium at high pressures, is to seed the hot hydrogen isotopes into expanding jets of helium, as in Ref.~\cite{huntington2023intense}. The detailed dynamics would then depend on energy transfer to the expansion gas via H-He and T-He scattering. In particular, the relative
efficacy of the system for the projected T application based on more
easily studied H test systems depends on the ratio of these cross sections. 
\end{enumerate}

In a different context, the dispersion of atom velocities also limits the ultimate precision available for the spectroscopic measurements of hydrogen isotopes. An intense source of cold  atomic hydrogen isotopes can improve the precision of 1S-2S spectroscopy, a crucial benchmark process for few body quantum electrodynamics. It also has promising applications in searches for gravitational quantum states~\cite{killian2024grasian}, new short-range forces, CPT and Lorentz invariance violation~\cite{kostelecky2015lorentz}, and could itself serve as a cold tritium source for neutrino mass searches.  This leads to an additional physics case where these cross sections are required,

\begin{enumerate}

\item[4] The GRASIAN collaboration is developing new techniques for production and trapping of cold hydrogen isotopes~\cite{semakin2025cryogenic} using cryogenic RF discharges. Buffer gas cooling of deuterium/tritium  on  helium vapor is an important ingredient in the scheme, necessary since scattering of the heavier isotopes on superfluid helium surfaces is not as effective for cooling as it is for hydrogen~\cite{Berkhout1993} due to the ease with which they penetrate into the helium film~\cite{Kurten1985, Saarela1993}. The energy-dependent, low temperature tritium-helium and deuterium-helium cross sections are key inputs to performance calculations that inform the design parameters for such a system~\cite{semakin2025cryogenic}, and they have not been previously available.
\end{enumerate}

The above cases motivate the present work, which provides not only tritium-helium scattering cross section predictions, but also those for all hydrogen isotopes on both helium isotopes.

\begin{figure}
\begin{centering}
\includegraphics[width=0.99\columnwidth]{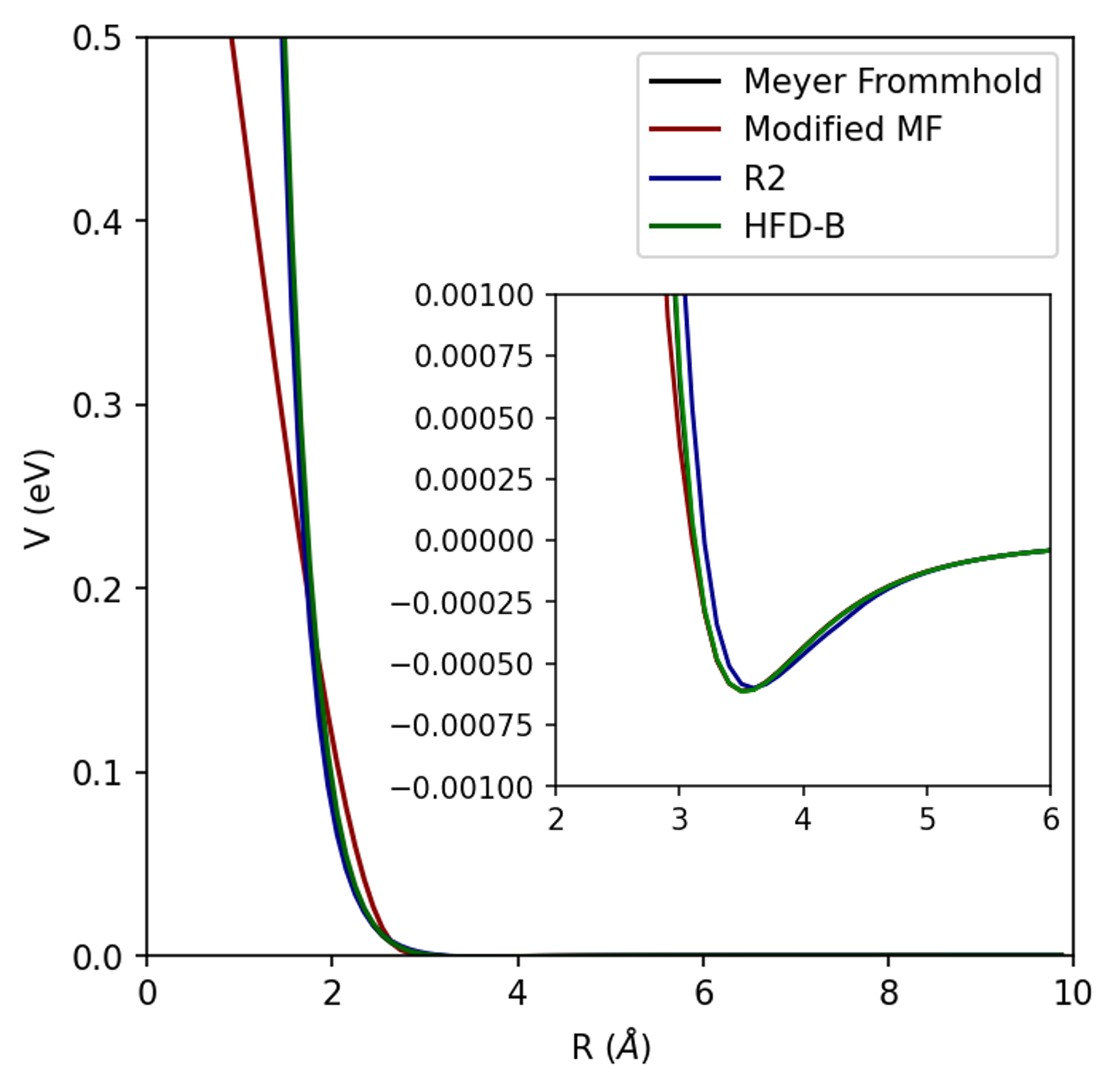}
\par\end{centering}
\caption{Potentials used in this work. The computationally derived Meyer Frommhold potential is digitized from Ref.~\cite{meyer1994long}, and the Modified MF potential includes the hard core modification proposed in Ref.~\cite{chung2002diffusion}. The R2 and and HFD-B potentials are experimentally motivated potentials as reported in Ref.~\cite{jochemsen1984diffusion}.}\label{fig:PotentialPlot}
\end{figure}

\section{Formalism and Potentials}

Because neither $^3$He nor $^4$He have any net electronic spin, it is notable that the scattering system considered here 
is considerably simpler than the T-T system of Ref~\cite{ElliottJones2025}. $^3$He has a nuclear spin, but the interaction
of nuclear spins is very weak compared with electronic spins due to the relative size of the electronic and nuclear magnetons. As such, the rate of inelastic / spin-changing interactions in the processes discussed here is negligible. The only relevant
process is thus elastic scattering.  Furthermore, because the two particles
involved in scattering are non-identical, no exchange
symmetry considerations must  be accounted for.   The elastic scattering cross section depends on center-of-mass
energy, reduced mass and inter-atom potential, but not on applied magnetic field, and can be obtained via a partial-wave phase-shift analysis. 

Following the approach outlined in Ref~\cite{ElliottJones2025}, and  similar to that employed in Ref.~\cite{Joudeh2013} for H-H elastic scattering and ~\cite{Al-Maaitah2012} for T-T elastic scattering, we solve the one-dimensional
Schrodinger equation for nuclear motion under the Born-Oppenheimer
approximation.  The relevant potential $V(R)$ is the ground state electron energy derived by solving for the three-electron ground state at fixed nuclear separation $R$, which is then used to model dynamical evolution of a nuclear coordinate wave function, $\psi(R)$.

Through axial symmetry only $m=0$ partial waves are relevant to the nuclear motion, and the radial wave function $u_{El}(R)$ for
each orbital angular momentum $l$ and energy $E$ satisfies the central
Schrodinger equation,
\begin{equation}
\left\{ \frac{d}{dR^{2}}+\frac{2\mu}{\hbar}\left[E-V(R)-\frac{l(l+1)\hbar^{2}}{2\mu R^{2}}\right]\right\} u_{El}(R)=0,\label{eq:Schrod}
\end{equation}
with $\mu$ the reduced mass and $\hbar$ Planck's constant. The large-distance
solution at $R\rightarrow\infty$ is expressed in terms of partial wave phase shifts $\delta_{l}$
as
\begin{eqnarray}
\lim_{R\to\infty}\psi(\boldsymbol{R})&=&e^{i\boldsymbol{kz}}+\frac{e^{ikR}}{R}f(\theta,\phi),\\
f(\theta)&=&\sum_{l=0}(2l+1)\frac{e^{(2i\delta_{l}(E)-1)}}{2ik}P_{l}(\cos(\theta)).
\end{eqnarray}
where $k$ is the wave vector.  The cross section can then be obtained from these phase shifts
as
\begin{equation}
\sigma=\int d\Omega|f(\theta,\phi)|^{2}=\frac{4\pi}{k^{2}}\sum_{l}(2l+1)\sin^{2}\left[\delta_{l}(E)\right].~\label{eq:XS}
\end{equation}
Eq~\ref{eq:XS} differs from the similar equation in of Ref.~\cite{ElliottJones2025} via 1) its pre-factor
of 4 rather than 8, and 2) and a sum over both odd and even values of $l$
rather than even-only, both of these adjustments deriving from the fact we are now handling scattering of distinguishable atoms rather than indistinguishable bosons. 

At low energy, the cross section is determined by the \textit{s}-wave scattering
length alone, per
\begin{equation}
\sigma=4\pi a_{s}^2,\\
\end{equation}
which is obtained from the low energy limit of the \textit{s}-wave phase shift $\delta_0(k)$ as
\begin{equation}
a_s=-\lim_{k\to0}\frac{\tan(\delta_{0}(k))}{k}.~\label{eq:ScatLength}
\end{equation}
All of the phase shifts are in practice found by solving Eq.~\ref{eq:Schrod} outward from
low $R$, and comparing the phase of the outgoing spherical wave at long-distance against the free-particle
solution of the same energy that is non-singular at the origin. This approach follows widely used methods in quantum scattering theory.~\cite{weinberg2015lectures}

To obtain scattering cross sections from Eq.~\ref{eq:Schrod} we require the appropriate scattering potential.  Several potentials are available for these calculations.  Some early H-He potentials were compiled by Jochemsen {\em et. al} in Ref~\cite{jochemsen1984diffusion}. Their R2 potential combines measurements of the short-range repulsive potential from molecular beam experiments~\cite{gengenbach1973determination} matched to a long-range dispersive part extracted from low energy elastic scattering~\cite{hardy1980magnetic,toennies1976determination} via an intermediate Lennard-Jones form, with parameters adjusted to match experimental data.  It is notable that this experimentally derived potential is based on data taken in the one-to-few kelvin regime, and so reliability at much lower energy remains an open question.  In the same work, another potential is presented, reported via private communication from Scoles and based on
self-consistent field calculations of Ref.~\cite{das1978calculated} matched to the long-range dispersion terms of Ref~\cite{tang1976upper}
connected via damping function. This computationally derived potential is also used in our calculations, denoted as in Ref.~\cite{jochemsen1984diffusion} as HFD-B.  

More recently, in Ref.~\cite{meyer1994long} Meyer and Frommhold computationally re-evaluated the interaction potential 
as a function of atomic separation in the H-He system, using a multi-reference
configuration interaction method. The Meyer Fromhold potential is
shown in Fig.~\ref{fig:PotentialPlot}.  Using the Meyer Frommhold potential, in Ref~\cite{chung2002diffusion} Chung and Dalgarno calculated the diffusion constants for hydrogen in helium
and helium in hydrogen. Comparison to experimental results at ambient
temperature and pressure showed a 15\% discrepancy, leading the authors
to consider a modified potential with a steeper repulsive core to better
match these data while maintaining agreement between the \textit{ab initio} potential and diffusion measurements at 1~K.   The steep
inner core difference between modified and un-modified Meyer Frommhold potentials will be shown to be most relevant for the lighter combinations at the lowest energies, but marginal over most of the parameter space considered for D-He and T-He scattering.

In this work we will consider the modified (mod-MF) and unmodified (MF) Meyer Frommhold potentials, as well as the Jochemsen R2 and Scoles HFD-B potential as inputs to momentum-dependent elastic scattering cross section calculations. All four potentials are shown overlaid in Fig.~\ref{fig:PotentialPlot}, with an inset showing a detailed view of the dip region that is especially relevant for low-energy scattering.

\begin{figure}
\begin{centering}
\includegraphics[width=0.99\columnwidth]{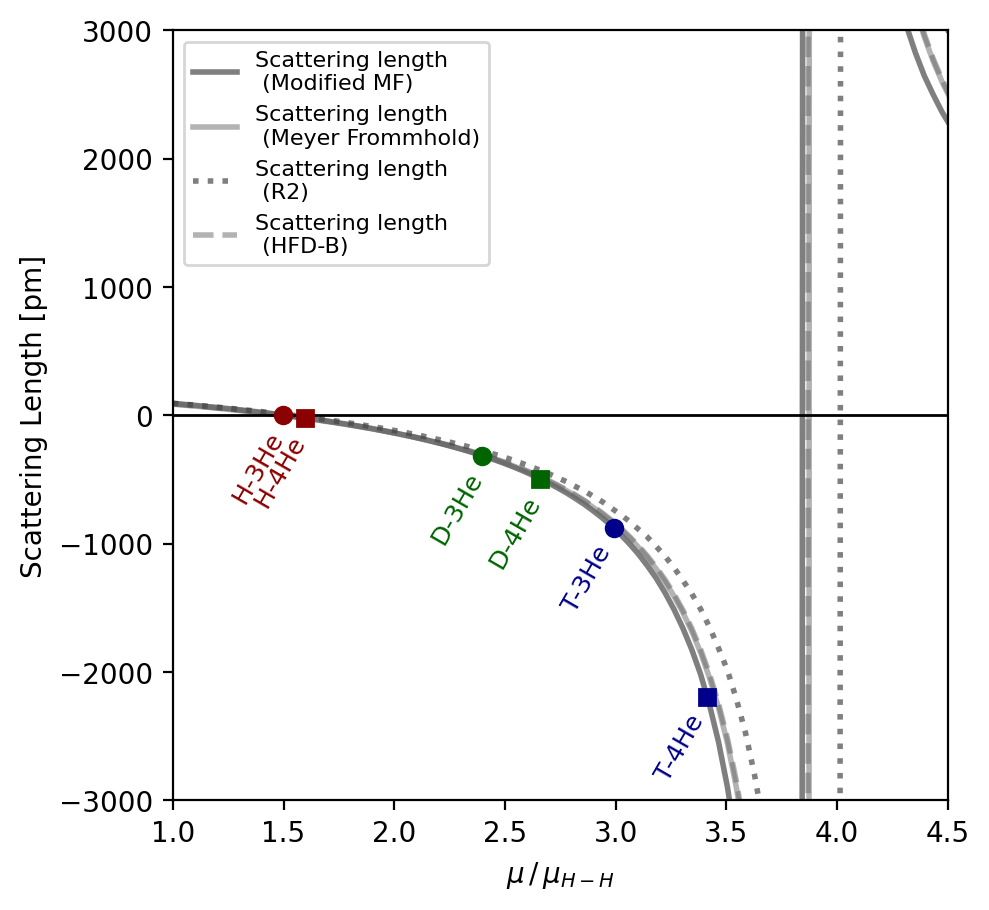}
\par\end{centering}
\caption{Reduced mass dependent \textit{s}-wave scattering length $a_s$ for the H-He system as a function of reduced mass $\mu$ in units of the H-H reduced mass $\mu_{H-H}$.  The low energy scattering lengths for the various isotope combinations of interest are indicated on the curve and reported in Table~\ref{tab:s-wave-scattering-lengths} }\label{fig:Reduced-mass-dependent}
\end{figure}

\section{Results}

We first calculate the low energy limit of the scattering cross section, which depends only on the \textit{s}-wave scattering length via Eq.~\ref{eq:ScatLength}. Since there are no non-trivial spin or magnetic-field dependencies, all of the
(H,D,T) on ($^{3}$He, $^{4}$He)
scattering cross sections at $k\rightarrow 0$ can be represented as a universal function of the reduced mass $\mu$ used in solution of Eq.~\ref{eq:Schrod}. Fig~\ref{fig:Reduced-mass-dependent} shows the predicted  \textit{s}-wave scattering length as a function of $\mu$ with the
experimentally relevant values marked, where the reduced masses are reported relative to the H-H scattering system $\mu_{H-H}=8.3401\times10^{-28}$~kg. Table~\ref{tab:s-wave-scattering-lengths}
provides the value of $\mu$ and the \textit{s}-wave scattering length that determines the low-temperature
cross section for each two-body processes. As an interesting point of comparison, Chung and Dalgarno reported
an \textit{s}-wave scattering length of $a_{s}=0.359$~Bohr for H-$^4$He using
the modified Meyer Frommhold potential, which compares favorably against our calculated value of $a_{s}=0.360$~Bohr, in these units.  This agreement provides a useful validation of our methodology and implementation.

\begin{table}
\begin{centering}
\begin{tabular}{|c|c||c|c|c|c|}
\cline{3-6}
\multicolumn{1}{c}{} & \multicolumn{1}{c|}{} & \multicolumn{4}{c|}{$a_{s}$ (pm)}   \tabularnewline
\hline
\bf Species &   $\mathbf{\mu/\mu_{H-H}}$ &{\em mod-MF} & {\em MF}& {\em R2}& {\em HFD-B} \tabularnewline
\hline 
\hline 
H-$^3$He & 1.499 & 3.85 & -0.69& 10.58& 1.49\tabularnewline
\hline 
H-$^4$He & 1.598 & -19.1 & -12.1&-10.31& -20.65\tabularnewline
\hline 
D-$^3$He & 2.396 & -312 & -309&-273& -305\tabularnewline
\hline 
D-$^4$He & 2.658 & -493 & -481&-426&-475\tabularnewline
\hline 
T-$^3$He & 2.993 & -880 & -842&-735 &-834\tabularnewline
\hline 
T-$^4$He & 3.414 & -2200 & -1990&-1614&-1965\tabularnewline
\hline 
\end{tabular}
\par\end{centering}
\caption{\textit{s}-wave scattering lengths (zero energy limit) for each isotope combination calculated using the modified Meyer Frommhold (mod-MF) and original Meyer Frommhold (MF) and Jochemsen (R2) and Scoles (HFD-B) potentials.  The low energy limit of the scattering cross section can be obtained from Eq.~\ref{eq:ScatLength}.}\label{tab:s-wave-scattering-lengths}

\end{table}

Because the potential is attractive at long distances, bound states can be supported as long as the atomic masses are sufficiently large to stabilize the van der Waals dimer.  The clear pole in Fig.~\ref{fig:Reduced-mass-dependent} shows that the first bound state is introduced ({\em i.e.}, becomes present with energy E$\sim$0) for reduced masses of $\sim$3.9 $\mu_{H-H}$.  This is higher in mass than the threshold for the similar bound state that was found to occur in the scattering of two hydrogen-like atoms, which appeared at $\sim$3.2 $\mu_{H-H}$~\cite{ElliottJones2025}.  The difference between these thresholds owes to the  difference between shapes of the H-He and H-H~\cite{Silvera1986} potentials.  Just as in the hydrogen-hydrogen scattering system, this near-resonant bound state leads to enhancements of the cross sections for heavier isotope combinations, with both T-$^3$He and T-$^4$He scattering being significantly stronger than the corresponding H-$^3$He and H-$^4$He scattering processes. As in T-T scattering, the enhancement factors are of order 10$^2$ in scattering length, corresponding to order 10$^4$ in cross section.

\begin{figure}[t]
\begin{centering}
\includegraphics[width=0.99\columnwidth]{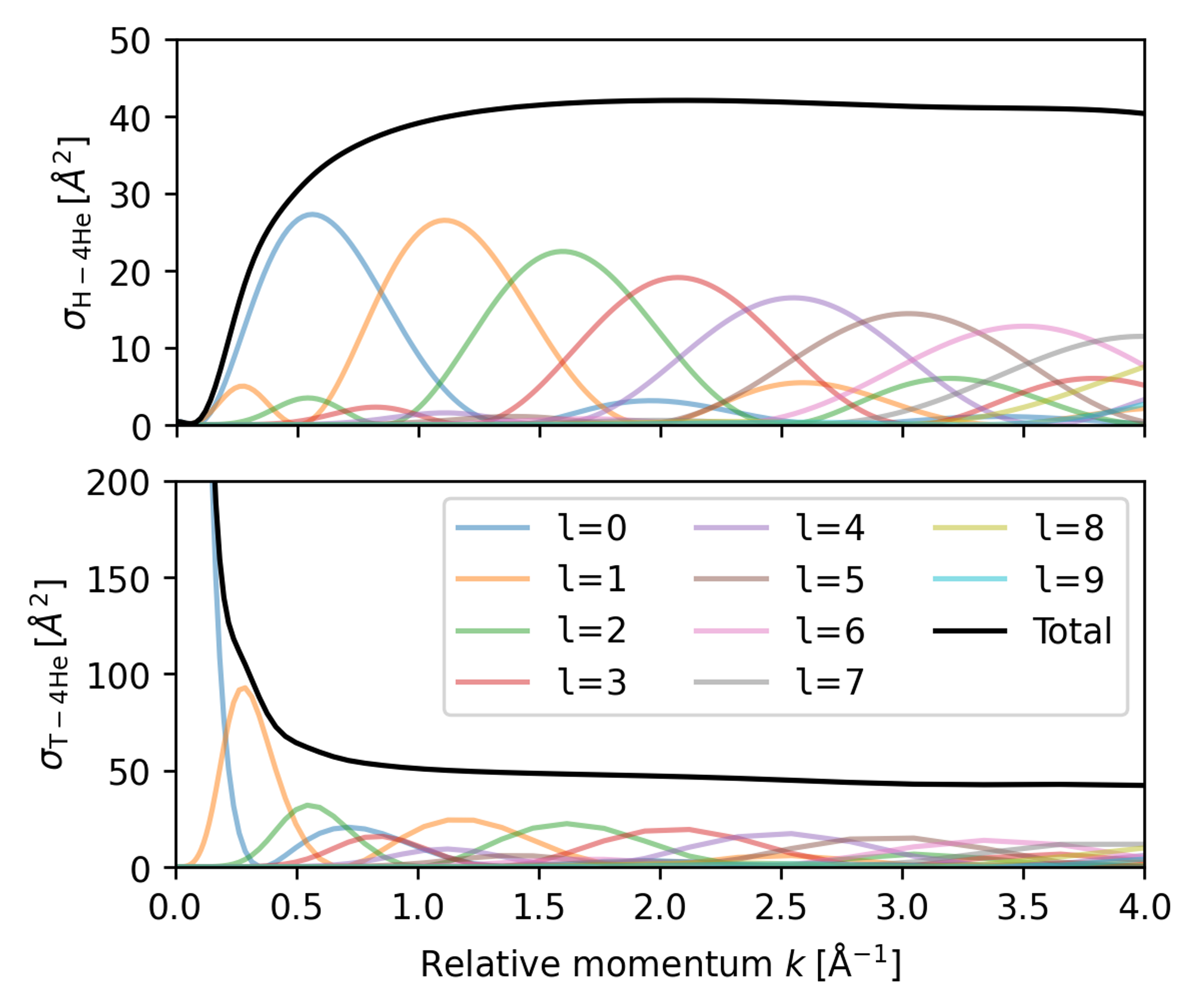}
\par\end{centering}

\caption{Partial wave components of the scattering cross sections for the H-$^4$He (top) and T-$^4$He (bottom) processes from the modified Meyer Frommhold potential. The results are similar over most relative momentum values, with the exception of the resonantly enhanced \textit{s}-wave amplitude at low energy, as shown on Fig.~\ref{fig:Reduced-mass-dependent}.\label{fig:PartialWaves}}

\end{figure}

\begin{figure}[t]
\begin{centering}
\includegraphics[width=0.99\columnwidth]{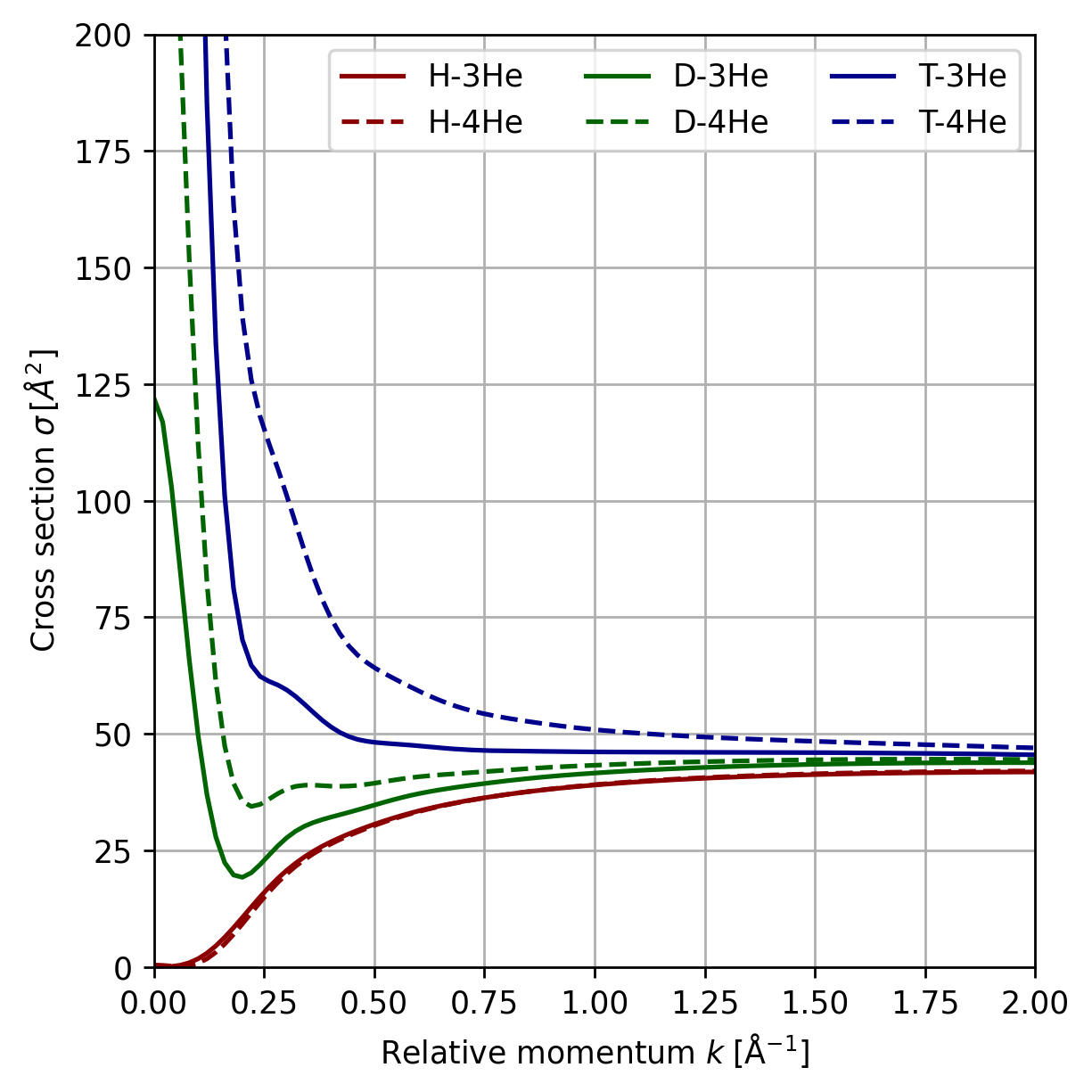}\\
\par\end{centering}

\caption{Momentum-dependent cross section on linear scale for each of the processes considered in this paper calculated using the modified Meyer Frommhold potential.  \label{fig:AllXSMom}}

\end{figure}

To calculate momentum-dependent cross sections we require the phase
shifts at non-zero energy and also the higher partial wave contributions.
Fig.~\ref{fig:PartialWaves} shows example partial wave decompositions of the H-$^{4}$He
and T-$^{4}$He processes, and Fig.~\ref{fig:AllXSMom} shows the summed cross section in each channel.  Because the hydrogen cross sections have an especially small $s$-wave scattering length, they increase with energy, first via a rising $s$-wave contribution and then through higher partial waves.  The heaver isotopes have larger scattering lengths due to the aforementioned pole at $\mu\sim$3.9 $\mu_{H-H}$, so their  cross sections fall with energy as the large $s$-wave contribution gives way to the un-enhanced, higher partial waves.  At the highest energies, the high-$l$ partial wave amplitudes superpose to provide a constant cross section, which becomes independent of both energy and reduced mass for hard collisions.   This is the geometric limit, where the process becomes an effectively hard-sphere scatter with the cross section determined by the atomic van der Waals radii.

\begin{figure}[t]
\begin{centering}
\includegraphics[width=0.99\columnwidth]{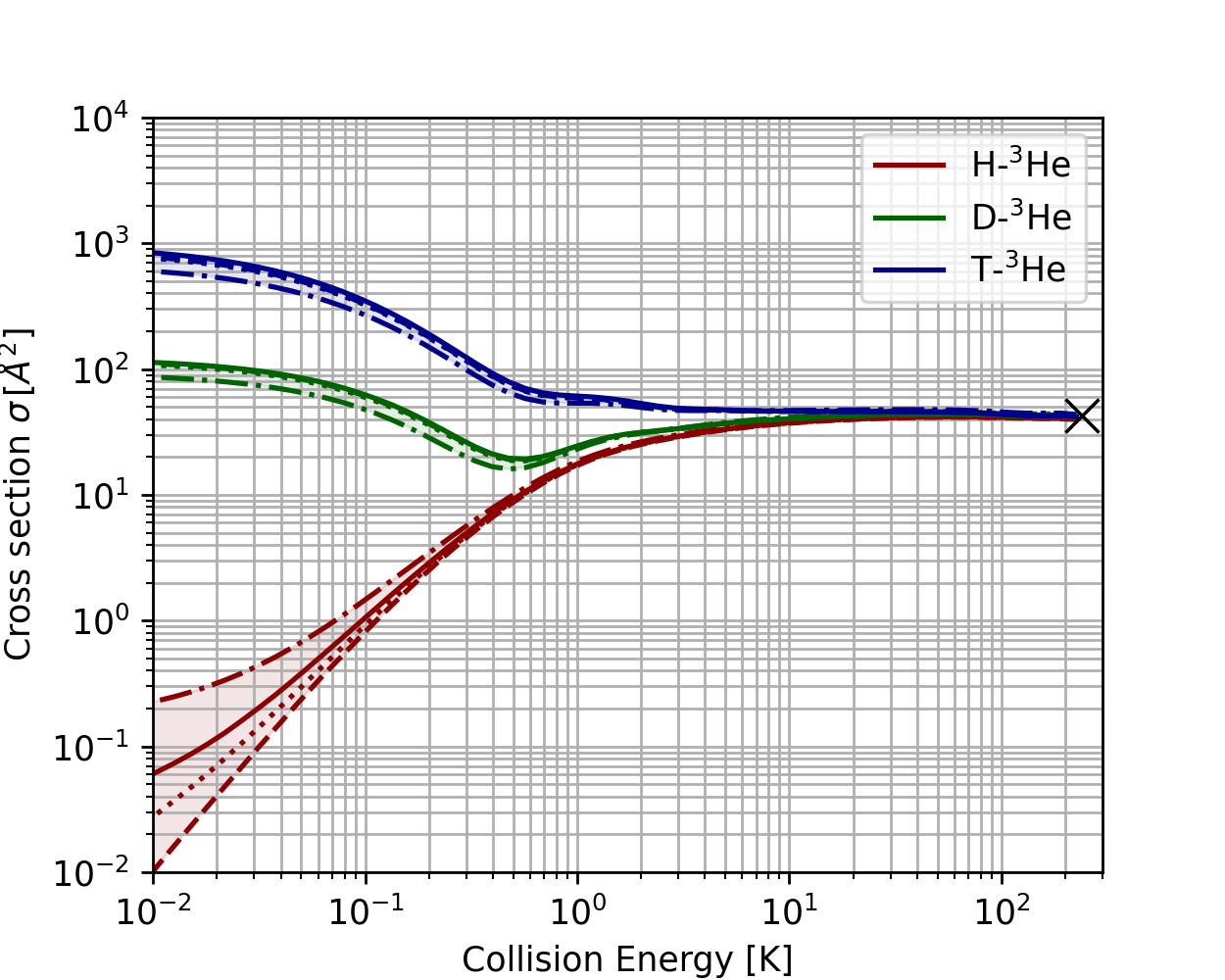}\\
\includegraphics[width=0.99\columnwidth]{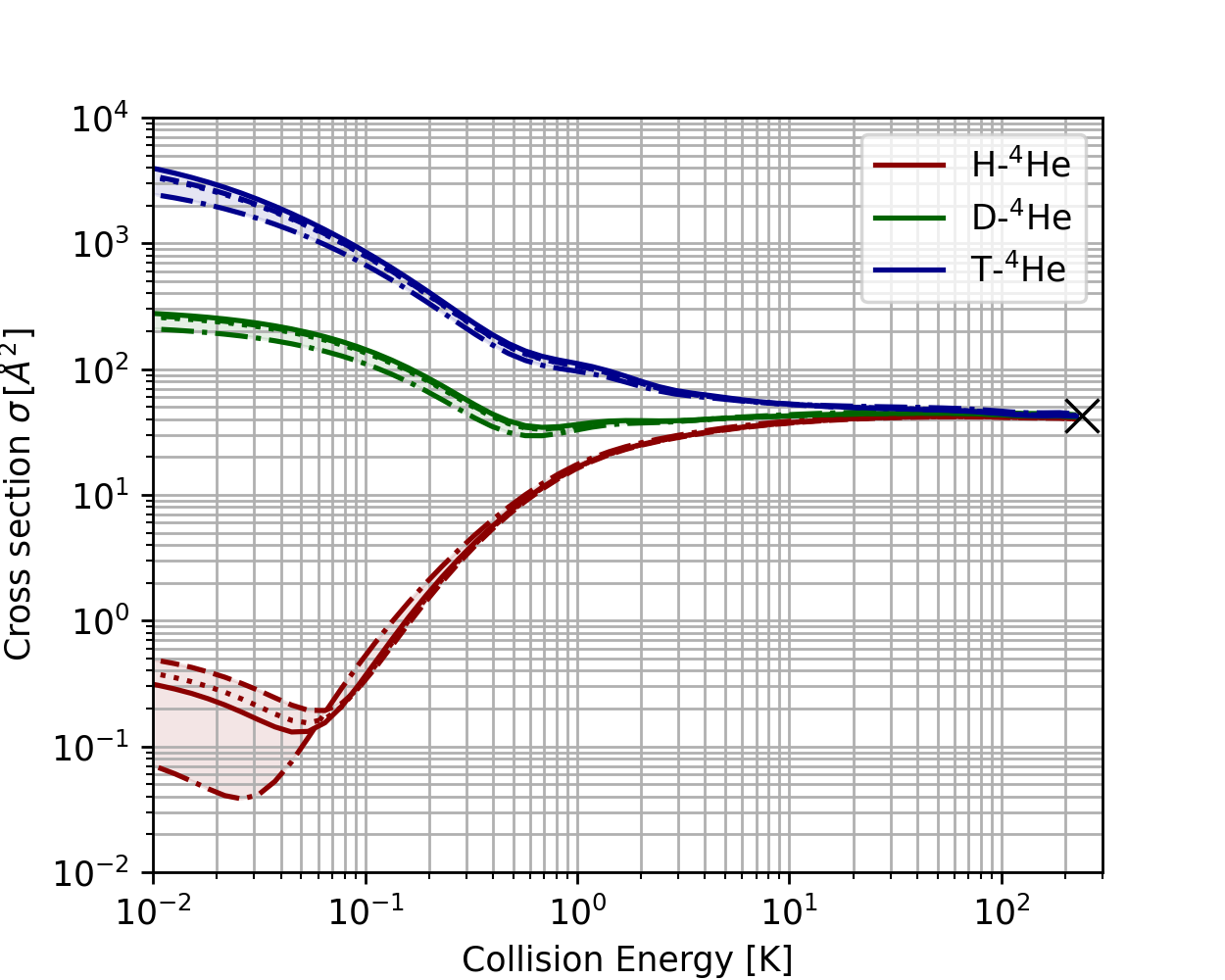}
\par\end{centering}

\caption{Energy-dependent cross section for each of the processes considered in this paper for $^3$He (top) and $^4$He (bottom). The energy reported is the center of mass collision energy in units of Kelvin,  $p^2/(2\mu k_B)$.  The black X shows the hard-sphere cross prediction based on the Van Der Waals radii of hydrogen and helium. Different line styles represent results obtained using different interaction potentials: MF (solid), MFmod (dashed), R2 (dash-dotted), and HFD-B (dotted).} \label{fig:AllXS}

\end{figure}

Fig.~\ref{fig:AllXS} shows the predicted energy-dependent cross sections for each of the isotope pairs over a wide range of temperatures, and is the primary result of this work.  As previously observed, the low energy cross sections are dictated by the \textit{s}-wave scattering length enhancement with a strong dependence on reduced mass of the colliding pair. The high energy cross sections converge to a common limiting value, as expected given the equivalent electronic structure in each case.  It is notable that this limit is consistent with the expected high energy hard-sphere scattering cross section of $\sigma=2\pi r^2$ (including the factor of 2 for shadowing in quantum scattering~\cite{sakurai2020modern}) with $r$ given by the summed van der Waals radii of hydrogen ($r_{H}=100$~pm) and helium ($r_{He}=140$~pm). This ``black disc limit'' reference point is marked with a cross at the high energy side of Fig.~\ref{fig:AllXS}.

For the majority of the atom pairs considered, the various potentials provide comparable
predictions, yielding a similar scattering length.   The exceptions are the low energy limits of the H-$^{3}$He and H-$^{4}$He processes, which are close in reduced mass to the zero-crossing of the scattering length vs $\mu$ curve crosses $a_{s}=0$ in Fig. \ref{fig:Reduced-mass-dependent}. In these channels, the various potentials still provide largely consistent results above approximately 100~mK.  This suggests their reliability at least an order of magnitude lower in temperature than the datasets from which they have been derived or  previously compared.  Below 100~mK, however, the predictions diverge significantly.  It is noted that although predictivity is limited in this very low energy regime, such collisions would be difficult to arrange in the majority of experimental configurations, since the helium ground state is not magnetically trappable and will have very low vapor pressure in any cell held at this temperature.   Nevertheless, it is cautioned that in any cases where they could be of experimental relevance, an accurate prediction of the low energy H-$^{3}$He and H-$^{4}$He cross section in particular would benefit from development of more precise ab-initio potentials.  

One other deviation between predictions from different potentials is apparent at low energy. The T-$^{4}$He process scattering length obtained using the Jochemsen R2 potential is separated from the other predictions, being significantly affected by the shift of the \textit{s}-wave resonance position in Fig~\ref{fig:Reduced-mass-dependent}.  Since the semi-empirical Jochemsen R2 potential has not been tuned or exercised in this very low energy range, we recommend to rely instead on the computationally derived Meyer Frommhold and Scoles potentials in this regime,  all of which give consistent predictions at the $\sim$5\% level~\cite{chung2002diffusion}.   

It is to be noted that in the most of the experimentally relevant parameter space, the differences between calculations made using any of the potentials are marginal.  All heavier isotope combinations find consistent predictions at the few percent level above 50~mK, with the H-$^3$He and H-$^4$He processes obtaining similar consistency above 200~mK.  This suggests a promising degree of predictivity for the applications that have motivated the present work.

\section{Conclusion}

We have reported calculations of the elastic scattering cross sections of H, D, and T on $^3$He and $^4$He.  Although considerably simpler than the calculations of T-T cross sections presented in Ref.~\cite{ElliottJones2025} due to the lack of spin-spin interactions, most of the presented cross sections have not been previously reported in the literature. Their values are required to inform aspects of the design and development of atomic sources for neutrino mass experiments and Doppler-free two-photon 1S–2S spectroscopy in atomic deuterium and tritium.  

We find that the tritium-on-helium scattering cross sections exhibit a similar resonant enhancement to tritium-on-tritium triplet elastic scattering, albeit with a different resonance threshold in reduced mass, owing to the differently shaped interaction potential.  The energy dependent cross sections are presented for all atom combinations, and show a wide spread of values at low energy due to mass dependent enhancement of the \textit{s}-wave contribution, but convergence to a common value that is consistent with the black disc limit at energies above a few tens of kelvin. 

The rising deuterium-on-helium and tritium-on-helium scattering cross sections at low energy appear particularly promising for buffer gas cooling of these isotopes, since they encourage better thermalization with cold helium buffer gas.  These large low temperature  cross sections also indicate improved  performance of supersonic expansion sources with seeded helium jets, which could also have important technological implications for development of cold tritium sources using these techniques.  

The complete energy-dependent cross sections  are supplied as tables in the supplementary information~\cite{DataTable}, and an open-source code is provided for reproducing these results~\cite{Code}.

\section*{Acknowledgments}

BJPJ and AN are supported by the US Department of Energy under awards DE-SC0024434 and DE-SC0019223, and the Gordon and Betty Moore Foundation, grant DOI 10.37807/GBMF13782.  AS is supported by the Jenny and Antti Wihuri foundation.  We thank Sergey Vasiliev, Morgan Elliott and Paul Harmston for highly valued feedback during the development of these calculations.  We also thank the Project 8, QTNM and GRASIAN collaborations for providing the motivation for our wider program of low-energy hydrogen and helium isotope cross section evaluations.

\bibliography{PaperBib}

\end{document}